%% file: main.tex
\documentclass[10pt,twocolumn,oneside,a4paper,final]{IEEEtran}
\usepackage[utf8]{inputenc}
\def\BibTeX{{\rm B\kern-.05em{\sc i\kern-.025em b}\kern-.08em
    T\kern-.1667em\lower.7ex\hbox{E}\kern-.125emX}}

\usepackage{graphicx}
\usepackage{color}
\usepackage{cite}
\usepackage{amsmath}
\usepackage[caption=false]{subfig}
\usepackage{amssymb}
\usepackage{comment}
\usepackage{mathtools}
\usepackage{tabulary}
\usepackage{acronym}
\usepackage{upgreek}
\usepackage{algorithm}
\usepackage{algpseudocode}
\usepackage{placeins}

\DeclareMathOperator*{\argmax}{arg\!\max}

\newcommand{\myvec}[1]{\boldsymbol{#1}}

\include{acronyms}

\title{Deep Contextual Bandits for Orchestrating Multi-User MISO Systems with Multiple RISs} 
\author{\IEEEauthorblockN{Kyriakos Stylianopoulos$^1$, George Alexandropoulos$^1$, \\Chongwen Huang$^2$, Chau Yuen$^3$, Mehdi Bennis$^4$, and M\'{e}rouane Debbah$^5$}\\
\IEEEauthorblockA{$^1$Department of Informatics and Telecommunications,
National and Kapodistrian University of Athens, Greece\\
$^2$College of Information Science and Electronic
Engineering, Zhejiang University, China\\
$^3$Engineering Product Development Pillar,
Singapore University of Technology and Design, Singapore\\
$^4$Centre for Wireless Communications, University of Oulu, Finland\\
$^5$Technology Innovation Institute, Abu Dhabi, United Arab Emirates\\
}
\thanks{This work has been supported by the EU H2020 RISE-6G project under grant number 10101701 and by MOE Tier 2 MOE-000168-01.}\vspace{-0.7cm}}

\begin{document}

\maketitle

\begin{abstract}
    The emergent technology of Reconfigurable Intelligent Surfaces (RISs) has the potential to transform wireless environments into controllable systems, through programmable propagation of information-bearing signals. Techniques stemming from the field of Deep Reinforcement Learning (DRL) have recently gained popularity in maximizing the sum-rate performance in multi-user communication systems empowered by RISs. Such approaches are commonly based on Markov Decision Processes (MDPs). In this paper, we instead investigate the sum-rate design problem under the scope of the Multi-Armed Bandits (MAB) setting, which is a relaxation of the MDP framework. Nevertheless, in many cases, the MAB formulation is more appropriate to the channel and system models under the assumptions typically made in the RIS literature. To this end, we propose a simpler DRL approach for orchestrating multiple metasurfaces in RIS-empowered multi-user Multiple-Input Single-Output (MISO) systems, which we numerically show to perform equally well with a state-of-the-art MDP-based approach, while being less demanding computationally.

\end{abstract}
\begin{IEEEkeywords}
Reconfigurable intelligent surfaces, deep reinforcement learning, multi-armed bandits, multi-user MISO.
\end{IEEEkeywords}

\input{Sections/1_Introduction}
\input{Sections/2_System_Model}
\input{Sections/3_Prerequisites}

\input{Sections/4_Proposed_Methodology}

\input{Sections/5_Numerical_Evaluation}

\input{Sections/6_Conclusion}


\FloatBarrier
\bibliographystyle{IEEEtran}
\bibliography{references}

\end{document}

%% file: acronyms.tex
\newacro{ANN}[ANN]{Artificial Neural Network}
\acrodefplural{ANN}[ANNs]{Artificial Neural Networks}

\newacro{RIS}[RIS]{Reconfigurable Intelligent Surface}
\acrodefplural{RIS}[RISs]{Reconfigurable Intelligent Surfaces}

\newacro{6G}[6G]{Sixth Generation}
\newacro{SNR}[SNR]{Signal-to-Noise Ratio}
\newacro{SINR}[SINR]{Signal-to-Interference-plus-Noise Ratio}
\newacro{TX}[TX]{Transmitter}

\newacro{RX}[RX]{Receiver}
\acrodefplural{RX}[RXs]{Receivers} 

\newacro{TVC}[TVC]{Time Varying Channel}
\acrodefplural{TVC}[TVCs]{Time Varying Channels}

\newacro{BS}[BS]{Base Station}
\newacro{UE}[UE]{User Equipment}
\acrodefplural{UE}[UEs]{Users Equipments}

\newacro{MSE}[MSE]{Mean Squared Error}

\newacro{MAC}[MAC]{Medium Access Control}

\newacro{NMSE}[NMSE]{Normalized Mean Squared Error}
\newacro{EM}[EM]{Electro-Magnetic}
\newacro{CSI}[CSI]{Channel State Information}
\newacro{MISO}[MISO]{Multiple-Input Single-Output}
\newacro{NOMA}[NOMA]{Non-Orthogonal Multiple Access}
\newacro{OFDM}[OFDM]{Orthogonal Frequency-Division Multiplexing}
\newacro{QoS}[QoS]{Quality of Service}

\newacro{UAV}[UAV]{Unmanned Autonomous Vehicle}
\acrodefplural{UAV}[UAVs]{Unmanned Autonomous Vehicles}

\newacro{IoT}[IoT]{Internet of Things}
\newacro{LOS}[LOS]{Line Of Sight}
\newacro{NLOS}[NLOS]{Non-Line Of Sight}

\newacro{AoA}[AoA]{Angle of Arrival}
\acrodefplural{AoA}[AoAs]{Angles of Arrival}

\newacro{AoD}[AoD]{Angle of Departure}
\acrodefplural{AoD}[AoDs]{Angles of Departure}

\newacro{AWGN}[AWGN]{Additive White Gaussian Noise}
\newacro{SL}[SL]{Supervised Learning}
\newacro{RL}[RL]{Reinforcement Learning}
\newacro{DRL}[DRL]{Deep Reinforcement Learning}

\newacro{MDP}[MDP]{Markov Decision Process}
\acrodefplural{MDP}[MDPs]{Markov Decision Processes} 

\newacro{POMDP}[POMDP]{Partially Observable Markov Decision Process}
\acrodefplural{POMDP}[POMDPs]{Partially Observable Markov Decision Processes}

\newacro{DQN}[DQN]{Deep Q Network}
\newacro{D$^3$QN}[D$^3$QN]{Duelling Double Deep Q Network}
\newacro{PG}[PG]{Policy Gradient}
\newacro{PPO}[PPO]{Proximal Policy Optimization}
\newacro{DDPG}[DDPG]{Deep Deterministic Policy Gradient}
\newacro{UCB}[UCB]{Upper Confidence Bound}
\newacro{GPU}[GPU]{Graphical Processing Unit}
\newacro{PDS}[PDS]{Post Decision State}
\newacro{TD}[TD]{Temporal Difference}
\newacro{MAML}[MAML]{Multi-Agent Reinforcement Learning}
\newacro{LSTM}[LSTM]{Long Short-Term Memory}

\newacro{EE}[EE]{Energy Efficiency}
\newacro{SE}[SE]{Spectral Efficiency}
\newacro{MIMO}[MIMO]{Multiple-Input Multiple-Output}

\newacro{IID}[IID]{Independent and Identically Distributed}

\newacro{MAB}[MAB]{Multi-Armed Bandits}
\newacro{CB}[CB]{Contextual Bandits}

\newacro{DRP}[DRP]{Deep Reward Prediction}

\newacro{DFT}[DFT]{Discrete Fourier Transform}
\newacro{ML}[ML]{Machine Learning}

%% file: Sections/1_Introduction.tex
\section{Introduction}\label{sec:introduction}
\vspace{-0.1cm}

The next era of wireless communication, i.e., the 6-th Generation (6G) networks, promises a plurality of remarkable benefits, such as orders of magnitude higher communication rates, ultra-low latency, sensing, and seamless integration of \ac{IoT} devices \cite{Saad_6G_2020}. Evidently, such an ambition necessitates the development of novel infrastructure and intelligent network components that can guarantee autonomous operation.
Among the candidate technologies, the suggestion of \acp{RIS} has been gaining momentum among academics, as well as major telecom vendors and operators. An \ac{RIS} is an artificial planar structure to be overlaid on the sides of unassuming surfaces, such as inner or outer walls of buildings \cite{EURASIP_RIS_paper,LZAWFM2020}. Its almost passive (i.e., without power amplification) metamaterials are organized in unit circuits that control the reflection angles of impinging \ac{EM} waves \cite{alexandg_2021}.
The deployment of an RIS is accompanied by a controller that: (a) controls the internal states of the \ac{RIS} elements, (b) communicates with other components of the environment, and (c) empowers RISs with computational capabilities, therefore yielding a dynamically programmable wireless environment ``as a service'' \cite{rise6g}.

To capitalize on the benefits of the RIS technology, however, and especially when considering the objective of high communication rates, the RIS controller incorporates some form of optimization procedure in order to select favorable states for the RIS elements. Traditional methods, derived from the broad field of optimization theory \cite{huang2019reconfigurable}, are in general excessively time-consuming to be deployed in real-time operations, and usually make strong assumptions about the properties of the underlying system. An alternative paradigm is advocated by the surging developments in \ac{ML}.
This domain of data-driven approaches involves a training process in which employed functions are fitted to observed data (e.g., from simulations or field trials) and are, thus, designed to be model-agnostic.
A lot of effort has been made toward adopting \ac{ML} approaches in wireless communications \cite{5GPPP_AIML}, and recently in \ac{RIS}-empowered systems \cite{AIRIS, huang2019spawc_indoor}. Their majority, however, considers methodologies that adhere to supervised learning, which is less equipped to deal with continual environmental changes, due to the inherent and separate data-collection process that takes place prior to the final deployment.

To make \acp{RIS} operate genuinely autonomously, online methods are required.
To that purpose, a growing area of research focuses on \ac{DRL} algorithms.
Most papers in the literature concern the sum-rate maximization problem by iteratively configuring the digital precoder and the RIS phase configuration \cite{Abdelrahman2020TowardsStandaloneOperation, Huang2020DRL_RIS, Feng2020DRL_MISO, Huang2021MultiHop, kim2021multiirsassisted}.
A number of works include further constraints and variations, such as controlling the power allocation \cite{Liu2021MNOMA_deployment} and deploying \acp{UAV} \cite{Samir2021AgeOfInformation}, while others prioritize different objectives, like secrecy rate \cite{Yang2021DRL_for_Secure}, energy efficiency \cite{Lee2020DRL_EE}, and resource scheduling \cite{gao2021ResourceAllocation, alhilo2021reconfigurable}.

A common trait in all relevant \ac{DRL} works is that they are based on the \ac{MDP} formalism, which constitutes the cornerstone of \ac{RL}.
In this article, we are motivated by the observation that the rate maximization problem, as it is typically framed in most works, constitutes a relaxed version of the \ac{MDP}, that nicely fits the elementary framework of \ac{MAB}.
Therefore, we propose a deep-learning-based bandit algorithm for the sum-rate maximization problem in RIS-empowered multi-user Multiple-Input Single-Output (MISO) systems, that is conceptually simpler.
Our evaluation process showcases that its performance is equal to a popular, state-of-the-art, \ac{DRL} algorithm, while having minimal hyper-parameters and lower neural network requirements.


\textit{Notation:} Bold-faced small and capital letters denote vectors and matrices, respectively. Calligraphy letters denote sets, unless specified otherwise. $[\mathbf{x}]_i$ denotes the $i$-th element of $\mathbf{x}$ and $[\mathbf{X}]_{i,:}$ ($[\mathbf{X}]_{:,i}$) denotes the $i$-th row (column) of $\mathbf{X}$. $\otimes$, ${\rm card}(\mathcal{S})$, and $x \sim A$ denote the Kronecker product, the cardinality of a set, and a random variable following a distribution, respectively. The ${\rm vec}(\cdot)$ operator vectorizes a matrix in row format, $\mathbb{E}\{ \cdot \}$ denotes an expectation, and $\mathbf{I}_{N}$ denotes the $N \times N$ ($N\geq2$) identity matrix. The complex standard Gaussian distribution is represented by $\mathcal{CN}(0,1)$.

%% file: Sections/2_System_Model.tex
\section{System Model and Design Objective}\label{sec:System_Model}

\subsection{System Model}\label{sec:System_specification}
\vspace{-0.1cm}
The considered downlink system consists of a \ac{BS}, equipped with $N_{\rm T}$ antenna elements, which serves $K$ single-antenna \acp{UE}. The direct link between the \ac{BS} and the \acp{UE} is assumed to be obstructed due to the presence of a blocker.
Instead, the communication is facilitated by $M$ identical \acp{RIS}, positioned at known locations to the BS.
Each surface is comprised of a planar arrangement of $N$ phase-shifting unit elements (organized in $N_{\rm h}$ rows and $N_{\rm v}$ columns).
Let $N_{\rm tot}\triangleq MN$ be the total number of elements of all deployed RISs, and assume an \ac{RIS} controller that is able to regulate the configuration of all the elements of the RISs. For computational purposes, it is usually convenient to assume the RIS elements to be controlled in groups of $N_{{\rm group}}$, so that elements within a group share the same configuration.
On that account, let $\hat{N}$ be the number of individually controllable groups referring to all $M$ RISs, i.e., $\hat{N} \triangleq  N_{\rm tot} / N_{{\rm group}}$.

For simplicity, we consider quantized \acp{RIS} with $1$-bit resolution (i.e., two possible phases per element), as is common practice in manufactured prototypes \cite{alexandg_2021}, in the ideal case of unit-amplitude reflection coefficients.
By denoting with $\myvec{\theta}_m$ the $N$-element vector that corresponds to the combined configuration of the elements of the $m$-th RIS with $m=1,2,\ldots,M$, its reflection coefficients are denoted as
\begin{equation}\label{eq:phase_shifts}
    \myvec{\phi}_m \triangleq \left[ \exp{(j \pi [\myvec{\theta}_m]_1)}, \dots, \exp{(j \pi [\myvec{\theta}_m]_N)} \right]^T.
\end{equation}
Proceeding, we make use of the free-space pathloss model $L(d) \triangleq 20\log_{10}(4\pi d\lambda^{-1})$
that represents the power loss factor (in dB) at a certain distance $d$ and for a wavelength $\lambda$ of the carrier frequency.
The involved links are modeled as frequency-flat fading channels that change independently after the elapse of the duration of the channel coherence time.
We use $\mathbf{H}_m \in \mathbb{C}^{N\times N_{\rm T}}$ and $\mathbf{g}_{m,k}\in\mathbb{C}^{1\times N}$ with $k=1,2,\ldots,K$ to denote the channel coefficients of the $m$-th RIS to the \ac{BS} and the $m$-th RIS to the $k$-th \ac{UE} links, respectively. The BS transmitter employs the precoding matrix $\mathbf{V} \in \mathbb{C}^{N_{\rm T} \times K}$ from a discrete codebook $\mathcal{V}$ to transmit a row vector of information symbols $\mathbf{q} \in \mathbb{C}^{K\times1}$.
Each $k$-th column of $\mathbf{V}$ represents the unit-norm precoding vector selected for the $k$-th UE. 
Hence, the transmitted signal is constructed as $\mathbf{x} \triangleq \mathbf{V}\mathbf{q}$, assuming equal power allocation among the UE signals so that the transmission power is constrained by the total power budget~$P$.

Using the above, the end-to-end channel is denoted as
\begin{equation}\label{eq:end-to-end-channel}
    \mathbf{b}_k \triangleq \sum_{m=1}^M \sqrt{L\left(d_m\right)L\left(d_{m,k}\right)}\mathbf{g}_{m,k} \mathbf{\Phi}_m \mathbf{H}_{m},
\end{equation}
where $d_m$ is the distance between the $m$-th RIS and the \ac{BS}, $d_{m,k}$ is the distance between the $m$-th RIS and the $k$-th \ac{UE}, and $\myvec{\Phi}_m$ is defined as the $N \times N$ diagonal matrix who has the elements of vector $\myvec{\phi}_m$ placed in its main diagonal.
Each of the $k$ \acp{UE} receives in baseband the following signal:
\begin{align}\label{eq:receive-signal-model}
    y_k = \mathbf{b}_k[\mathbf{V}]_{:,k}[\mathbf{q}]_k+\sum_{i=1,\,i \neq k}^{K} \mathbf{b}_k[\mathbf{V}]_{:,i}[\mathbf{q}]_i+ n_k,
\end{align}
where $n_k\sim\mathcal{CN}(0,\sigma^2)$ is the \ac{AWGN} corresponding to the $k$-th UE.

\vspace{-0.3cm}
\subsection{Channel Model}
\vspace{-0.1cm}
The Ricean fading model is used to characterize the channel gain matrices. Each of $\mathbf{H}_m$ and $\mathbf{g}_{m,k}$ consists of a mixture of a deterministic \ac{LOS} and a stochastic \ac{NLOS} components.
Specifically, the \ac{BS}-{RIS} links can be mathematically expressed as
\begin{equation}\label{Eq:H_m}
    \mathbf{H}_m \triangleq  \underbrace{\sqrt{\frac{\kappa_1}{\kappa_1+1}} \mathbf{\bar{H}}_m}_{\text{LOS component}} +  \underbrace{\sqrt{\frac{1}{\kappa_1+1}} \mathbf{\tilde{H}}_m}_{\text{NLOS component}},
\end{equation}
where $[\mathbf{\tilde{H}}_m]_{i,j}\sim\mathcal{CN}(0,1)$ and the LOS component $\mathbf{\bar{H}}_m$ is expressed in terms of the steering vector for the rectangular \ac{RIS} with ideal isotropic elements \cite{Alkhateeb_JSTSP_all} and the steering vector of the \ac{BS}.
Both of them depend on the azimuth and elevation angles of arrival and departure of the impinging/outgoing signals. Similarly, each \ac{RIS} to the $k$-th \ac{UE} link is modeled as
\begin{equation}\label{eq:g}
    \mathbf{g}_m \triangleq  \underbrace{\sqrt{\frac{\kappa_2}{\kappa_2+1}} \mathbf{\bar{g}}_{m,k}}_{\text{LOS component}} +  \underbrace{\sqrt{\frac{1}{\kappa_2+2}} \mathbf{\tilde{g}}_{m,k}}_{\text{NLOS component}},
\end{equation}
with the random vector $\mathbf{\tilde{g}}_{m,k} \sim \mathcal{CN}(0,\mathbf{I}_K)$ and the steering vector component $\mathbf{\bar{g}}_{m,k}$ depending on the relative positions between the $m$-th \ac{RIS} and the $k$-th \ac{UE}.
The Ricean factors $\kappa_1$ and $\kappa_2$ control the LOS-dominance of each channel.

\vspace{-0.3cm}
\subsection{Design Problem Formulation}
\vspace{-0.1cm}
As is common in multi-user communications, the \ac{SINR} metric is employed to describe the quality of transmissions.
Assuming that \ac{CSI} measurements can be obtained precisely and efficiently during a dedicated phase \cite{LZAWFM2020,Lin2021,Alamzadeh2021ris}, the \ac{SINR} for each $k$-th UE is computed as
\begin{equation}\label{eq:sinr}
{\rm SINR}_k \triangleq \frac{\left| \mathbf{b}_k[\mathbf{V}]_{:,k}\right|^2}
    {\sum_{i=1,\,i \neq k}^{K}  \left| \mathbf{b}_k [\mathbf{V}]_{:,i}\right|^2+\frac{K\sigma^2}{P}},
\end{equation}
which can be used for calculating the sum-rate performance in bits per second per Hertz as follows:
\begin{equation}\label{eq:rate}
    R_k \triangleq \log_2{\left(1 + {\rm SINR}_k \right)}.
\end{equation}

Given a time horizon $T$ of \ac{IID} channel realizations measured at discrete time intervals, the problem considered in this paper is that of the maximization of the sum rate among all UEs for the specified period. The free parameters of the system are the selection of the precoding matrix and the joint configuration of the $M$ \acp{RIS}.
A centralized controller is conceived, that observes at every time step all involved channel coefficients, selects the appropriate precoder and RIS configurations, which are then used for the signal transmission. For notation purposes, we incorporate the configurations of all the individually controlled RIS element groups in the binary $\hat{N}$-dimension vector $\mathbf{\vartheta}$.
The design optimization problem can now be summarized as:
\begin{align}
    \mathcal{OP} :& \underset{\mathbf{V} \in \mathcal{V}, \mathbf{\vartheta} \in \{0,1\}^{\hat{N}}}{\max} \sum\limits_{t=1}^{T}\sum\limits_{k=1}^{K}R_k 
    \quad \text{s.t.}\quad \mathbb{E}\{\|\mathbf{x}\|^2\}\leq P. 
\end{align}
The constraint needs to be satisfied by the design of the precoder and the power allocation onto the UE symbols. Nevertheless, $\mathcal{OP}$ is a discrete optimization problem of high-dimensionality, which typically involves applying an iterative optimization scheme at every channel coherence time.

%% file: Sections/3_Prerequisites.tex
\section{DRL-Based Problem Formulation }\label{sec:prerequisites}
\vspace{-0.1cm}
RL is a sequential decision making framework under which, at each discrete time step $t$, an agent (which is the controller in our RIS-empowered multi-user MISO system), observes the state of the environment (the wireless system) and decides on an action. The action is then transmitted to the environment, which feeds a reward signal back to the agent and proceeds to the next time step $t+1$. In the following, we give the correspondence of these concepts to the problem at hand:
\begin{itemize}
    
    \item \textbf{State:} Under the assumption of available and perfect \ac{CSI} at the agent, the state corresponds to the concatenated vector of all involved channel coefficients:
    \begin{align}\label{eq:state}
        \myvec{s}_t \triangleq [& {\rm vec}(\mathbf{H}_1), {\rm vec}(\mathbf{H}_2),\dots, {\rm vec}(\mathbf{H}_M), \nonumber \\
        &\mathbf{g}_{1,1}^T, \mathbf{g}_{1,2}^T,\dots, \mathbf{g}_{1,K}^T, \dots,  \\
        & \mathbf{g}_{M,1}^T,\mathbf{g}_{M,2}^T, \dots, \mathbf{g}_{M,K}^T ]^T.\nonumber
    \end{align}
    We denote by $\mathcal{S}$ the state space, which is a subset of $\mathbb{C}^{\rm dim}$, where ${\rm dim} 
    \triangleq M N (N_T + K)$.
    
    \item \textbf{Action:} The agent selects the precoding matrix and the joint RIS configuration, i.e., it is responsible to compute:
    \begin{equation}\label{eq:action}
        \myvec{a}_t \triangleq [ {\rm vec}(\mathbf{V}), \mathbf{\vartheta}^T]^T.
    \end{equation}
    In this work, we assume a discrete action space $\mathcal{A}$ with an implied ordering of the available actions. We will be using the notation $\mathbb{I}({\myvec{a}})$ to refer to the index of $\myvec{a}$ in $\mathcal{A}$.
    
    \item \textbf{Reward:} The reward $r_t$ is simply defined as the achievable sum-rate performance for the current \ac{CSI}, i.e.:
    \begin{equation}\label{eq:reward}
        r_t \triangleq \sum\limits_{k=1}^{K}R_k.
    \end{equation}
\end{itemize}

\begin{figure}[t]
    \centering
    \includegraphics[width=0.8\linewidth]{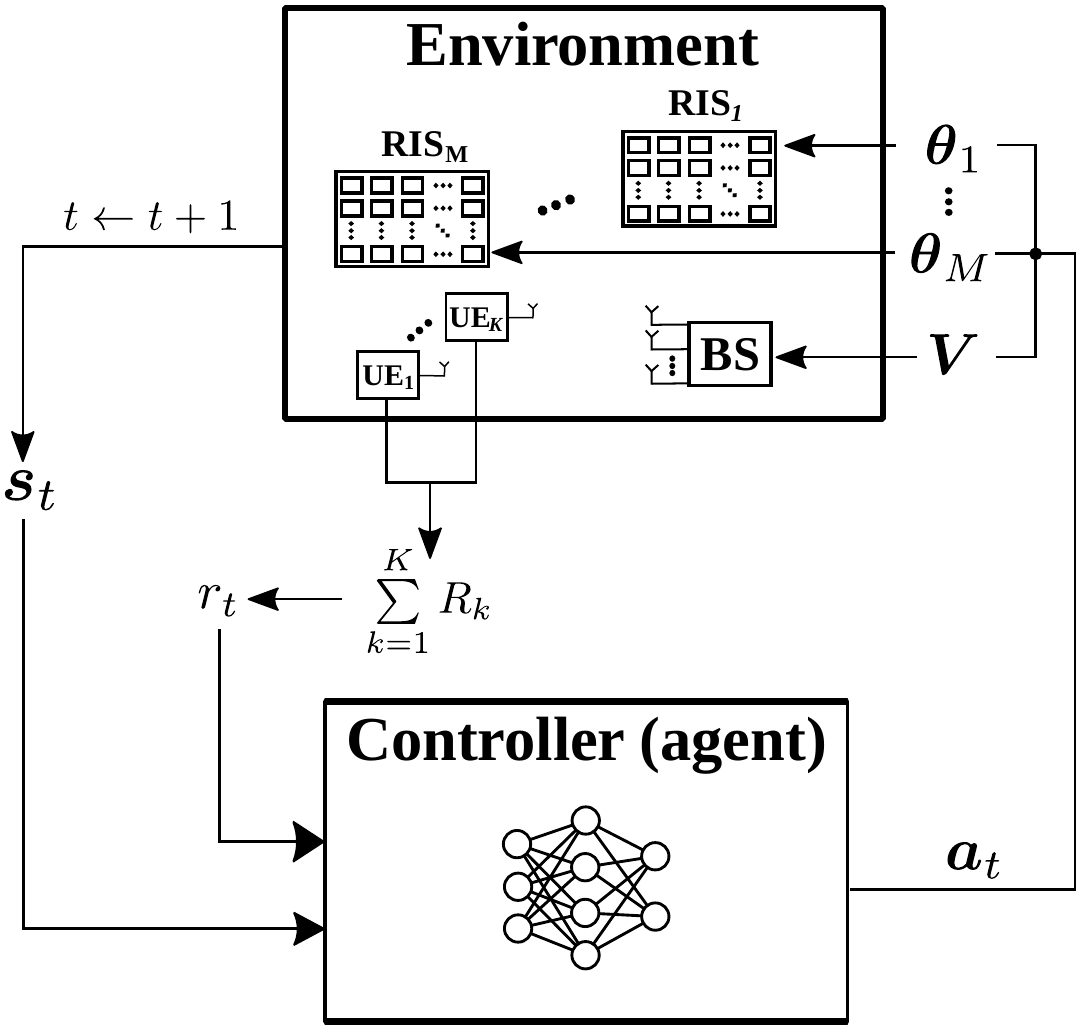}
    \vspace*{-7pt}
    \caption{An RL-based formulation of the sum-rate maximization problem in RIS-empowered multi-user MISO systems. The agent utilizes a neural network to learn from past experiences.}
    \label{fig:RL_RIS_formulation}
\end{figure}

It is assumed that one time step corresponds to one channel coherence block, hence, each state contains a different channel realization. The goal of the agent is to converge to a policy, i.e., a sequential action-selection function, that maximizes the (expected) sum of rewards during the interaction period. Hence, the objective of the \ac{RL} formulation is equivalent to $\mathcal{OP}$'s objective.
A schematic overview of the \ac{RL} process for the considered optimization problem is illustrated in Fig.~\ref{fig:RL_RIS_formulation}.

Problems that fall under the domain of \ac{RL} typically adhere to the \ac{MDP} formalism.
An \ac{MDP} is defined via the aforementioned state, action, and reward conceptualizations, but with the additional ``Markovian property'' of the environment dynamics, namely:
\begin{enumerate}
    \item The next state $\myvec{s}_{t+1}$ is produced by the environment according to a transition probability distribution (usually unknown to the agent), $\mathcal{P}$, that depends exclusively on the past state and action, i.e., $\myvec{s}_{t+1} \sim \mathcal{P}(\myvec{s}_{t+1}|\myvec{s}_t, \myvec{a}_t)$.
    \item The instantaneous reward at time $t$ is treated as a function that depends on $\myvec{s}_t$, $\myvec{a}_t$, and $\myvec{s}_{t+1}$, i.e., $r_t = \mathcal{R}(\myvec{s}_t, \myvec{a}_t, \myvec{s}_{t+1})$.
\end{enumerate}

In the growing field of \ac{DRL}, deep neural networks are employed to parameterize directly or indirectly the policy function, and are trained on collected experiences.
The main difference with other forms of ML is that the networks themselves guide the data collection process, instead of relying on a pre-compiled dataset, since they dictate the agent's interactions. As a result, there is an implicit trade off between the \textit{exploration} (observation of different states/actions) and the \textit{exploitation} (selection of already discovered beneficial actions) during the learning process.

The most popular algorithm in discrete action spaces is termed \ac{DQN} \cite{DQN}.
This agent is tasked with learning the optimal action value function $Q(\myvec{s}_t, \myvec{a}_t)$, that describes the expected ``utility'' (sum of rewards), when the agent observes $\myvec{s}_t$ and selects $\myvec{a}_t$.
\ac{DQN} chooses to approximate $Q$ with a neural network $Q_{\myvec{w}}$ with weights $\myvec{w}$ that receives a state as input and outputs a vector of dimension ${\rm card}(\mathcal{A})$, so that the $i$-th component of the vector is an estimate of the $Q$ value for the action with index $\mathbb{I}(\myvec{a}) = i$.
Once the $Q$ values are estimated, the agent's policy is to select the one with the highest $Q$ value, although a random action is selected with probability $\epsilon$ to encourage exploration (this behavior is commonly termed ``$\epsilon$-greedy''). The network can converge to the optimal $Q$ function by (partially) minimizing the squared \ac{TD} learning loss function of the network at every iteration, which is defined as:
\begin{equation}\label{eq:DQN-loss}
\begin{split}
    \mathcal{L}(\myvec{w}) \triangleq \sum\limits_{(\myvec{s}, \myvec{a}, r, \myvec{s}') \in \mathcal{B}} \left( Q_{\myvec{w}}(\myvec{s}, \myvec{a}) - (r + \underset{\myvec{a}' \in \mathcal{A}}{\max} Q_{\myvec{w}}(\myvec{s}', \myvec{a}')) \right)^2,
\end{split}    
\end{equation}
where $\mathcal{B}$ is a batch of collected past experiences (state, action, reward, and next state tuples).
The network's weights are updated at every step through any variant of gradient descent, using a learning rate $\eta_1$.
This process, however, is prone to instabilities during training. To that end, many variations were proposed that are usually applied in unison. Firstly, the collected experiences may be sampled in proportion to the resulting \ac{TD} error during the agent's last encounter. In addition, the gradient values may be clipped in a range $[-\delta, \delta]$.

More importantly, a ``target network'' $\hat{Q}_{\myvec{\hat{w}}}$ is introduced, which is a copy of the original $Q$ network, but with its own set of weights $\myvec{\hat{w}}$.
Its role is to be updated at a lower rate to help with the stability of the descent, as the original network changes.
The policy of \ac{DQN} now involves the selection of the action that maximizes $\hat{Q}_{\myvec{\hat{w}}}(\myvec{s}, \myvec{a})$, instead of $Q_{\myvec{w}}(\myvec{s}, \myvec{a})$, while $Q$ in the $\max$ term of \eqref{eq:DQN-loss} is also substituted by $\hat{Q}$.
While $Q$ continues to get updated via gradient descent at every time step $t$, the weights of $\hat{Q}$ change at a lower frequency $t'$ through a soft-copy from $\myvec{w}$, with a controllable ``temperature'' hyper-parameter $\tau$ as $\myvec{\hat{w}} \gets (1-\tau)\myvec{\hat{w}} + \tau \myvec{w}$.

%% file: Sections/4_Proposed_Methodology.tex
\section{Proposed MAB Methodology}\label{sec:methodology}
\vspace{-0.1cm}
Having described the theoretical aspects of the methodology and the benchmark algorithm in detail, we proceed with the presentation of our own contributions.
The main motivation for this work is the observation that (a) the channel realizations in the problem at hand are \ac{IID} and (b) the agent's action (RIS phase profiles and the BS precoder selection) result to the immediate calculation of the reward value (sum-rate), within the current coherent block. Under the prism of this inspection, it becomes apparent that the ``Markovian property,'' as defined in Section~\ref{sec:prerequisites}, is reduced to the degenerate case where $\mathcal{P}(\myvec{s}_{t+1}|\myvec{s}_t, \myvec{a}_t) \equiv \mathcal{P}(\myvec{s}_{t+1})$ and for the reward holds $\mathcal{R}(\myvec{s}_t, \myvec{a}_t, \myvec{s}_{t+1}) \equiv \mathcal{R}(\myvec{s}_t, \myvec{a}_t)$ (i.e., it is a function of only $\myvec{s}_t$ and $\myvec{a}_t$).
This has the profound effect of making the time steps disentangled from each other, in the sense that the agent's action cannot influence the environment's future evolution, and hence, move to more favorable states.
An agent is simply required to act greedily, by only considering the immediate reward, instead of devising a more sophisticated policy, usually attained by \ac{DRL} algorithms. 

This perspective motivates us to propose a conceptually simpler \ac{MAB}-based formulation for solving the sum-rate maximization design objective. In a \ac{MAB} setting, each available action is associated with an underlying distribution over the rewards, which is represented as $\mathcal{\hat{R}}(r|\myvec{a})$.
At each time step $t$, when an action $\myvec{a}_t$ is selected, a realization of the sum rate, $r_t$, is sampled from the distribution $\mathcal{\hat{R}}(r|\myvec{a}_t)$. Conceptually, \ac{MAB} algorithms keep track of running averages of the reward per action and use an exploration behavior for efficient search in the action space, such as the $\epsilon$-greedy strategy of \ac{DQN}.
The \ac{MAB} setting is able to admit ``context'' observations that may guide the agent toward the appropriate action selection at every time step, a variation denoted as \ac{CB}.
Note that in principle, no assumption is made about the generative process of the observations, other than that they influence $\mathcal{\hat{R}}(r|\myvec{a})$, i.e., $\mathcal{\hat{R}}(r|\myvec{a}) \equiv \mathcal{\hat{R}}(r|\myvec{s},\myvec{a})$.

\begin{figure}[t]
    \centering
    \includegraphics[width=0.9\linewidth]{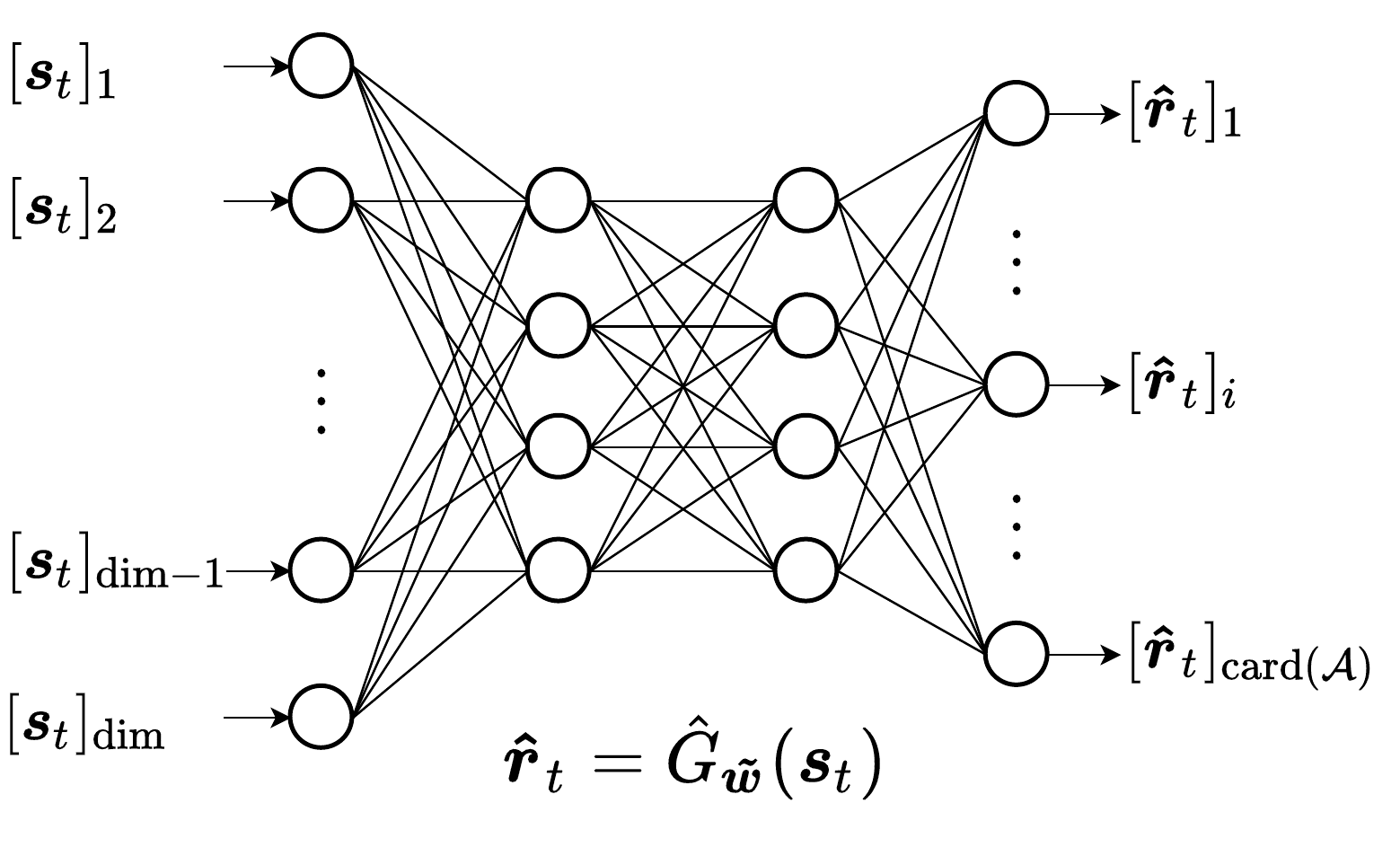}
    \vspace*{-10pt}
    \caption{The structure of the reward-prediction network  $\hat{G}_{\myvec{\tilde{w}}}$ of the proposed \acs{DRP} algorithm. The network receives a channel observation as input and outputs its predicted expected reward for each of the actions in the action space.}
    \label{fig:reward-network}
\end{figure}

One straightforward technique one may employ to solve the \ac{CB} problem is to have a neural network predicting the average reward for each action.
We propose one such bandit algorithm, which we call \ac{DRP} for the considered sum-rate maximization problem.
A neural network $\hat{G}_{\myvec{\tilde{w}}}(\myvec{s})$, parameterized by its weight vector ${\myvec{\tilde{w}}}$, receives as input a state/observation vector $\myvec{s}_t$ and outputs a vector $\myvec{\hat{r}}_t \in \mathbb{R}^{{\rm card}(\mathcal{A})}$, so that its $i$-th element corresponds to the network's prediction for the expected reward if the action with $\mathbb{I}(\mathbf{a})=i$ were to be selected upon observing $\myvec{s}_t$.
A schematic overview of this reward-prediction network is given in Fig.~\ref{fig:reward-network}.
Notice that the $\hat{G}$ network is very similar to the $Q$ network of DQN, however, their difference lies in the interpretation of the predictions (expected rewards versus $Q$-values) and the training process.
In fact, given that the transitions are \ac{IID} and unaffected by the agent's actions, the simple \ac{MSE} loss function is exploited to make the network accurately predict the expected values:
\begin{equation}\label{eq:loss-neural-bandit}
    \hat{\mathcal{L}}(\myvec{\tilde{w}}) \triangleq \left( r_t - [\hat{G}_{\myvec{\tilde{w}}}(\myvec{s}_t)]_{\mathbb{I}(\myvec{a}_t)} \right)^2 .
\end{equation}
In the above, the $[\hat{G}_{\myvec{\tilde{w}}}(\myvec{s}_t)]_{\mathbb{I}(\myvec{a}_t)}$ can be interpreted as a mask that only considers the reward prediction that corresponds to the actually selected action during the interaction.
Since the output of  $\hat{G}$ provides an estimate of how good each action is, the $\epsilon$-greedy strategy is utilized to help the agent explore different actions during training.
The complete proposed method is summarized in Algorithm~\ref{alg:neural-bandit}.
Note that, in contrast to the \ac{DQN} benchmark, our method requires only two hyper-parameters ($\epsilon$ and $\eta_2$) and a single instance of a neural network. 

\begin{algorithm}[!t]
\caption{The Proposed \acf{DRP}}
\label{alg:neural-bandit}
\begin{algorithmic}[1]
\Require Probability of selecting a random action $\epsilon$ and the learning rate $\eta_2$.
\State Initialize $\myvec{\tilde{w}}$ randomly.
\State Observe initial state $\myvec{s}_1$ from the environment.
\For{$t=1,2,\dots,T$}
    \State With probability $\epsilon$ select a random action from $\mathcal{A}$, 
    
    \hspace{-0.185cm}otherwise, select $\myvec{a}_t$ so that $\mathbb{I}(\myvec{a}_t) =  \argmax \ \hat{G}_{\myvec{\tilde{w}}}(\myvec{s}_t)$.
    \State Feed $\myvec{a}_t$ to the environment and receive $r_t$ and $\myvec{s}_{t+1}$.
    \State Compute gradient $\nabla \hat{\mathcal{L}}(\myvec{\tilde{w}})$ using \eqref{eq:loss-neural-bandit}.
    \State Update network as  $\myvec{\tilde{w}} \gets \myvec{\tilde{w}} - \eta_2 \nabla \mathcal{L}(\myvec{\tilde{w}})$.
\EndFor
\State \Return Trained network $\hat{G}_{\myvec{\tilde{w}}}$.
\end{algorithmic}
\end{algorithm}

%% file: Sections/5_Numerical_Evaluation.tex
\section{Numerical Evaluation}\label{sec:Numerical_Evaluation}
\vspace{-0.1cm}
To assess the performance of our proposed methodology, we devise a scenario with $K=2$ \acp{UE} and $M=2$ \acp{RIS} in the presence of \ac{LOS}-dominated channels.
The main parameters of the simulated RIS-empowered communications are given in Table~\ref{tab:setup-parameters}, although we allow for the total power budget $P$ and the number of \ac{RIS} elements $N$ (and subsequently $N_{\rm tot}$ and $\hat{N}$) to vary across the following evaluation settings.
The precoding codebook was constructed via the $2 \times 2$ \ac{DFT} matrix. In detail, its first two columns were considered to be the available choices for the precoder intended for the first \ac{UE}, while the latter two were allocated to the second \ac{UE}.
The codebook is purposely kept modest in order to restrain the exponential growth of the action space, in the view of investigating the effects of the \acp{RIS} in the considered communication system.

In our evaluation process, we consider the proposed bandit algorithm \ac{DRP} along with the \ac{DQN} benchmark.
We also simulate the classic \ac{UCB} \ac{MAB} algorithm \cite{UCB} that disregards any observations.
Instead, it keeps track of running averages and confidence intervals for the expected sum rates per action, and selects the one with the highest confidence bound.
Finally, the random action selection policy and the optimal policy of exhaustively evaluating all \ac{RIS} configurations and precoders at every channel realization are included as a baseline and upper bound, respectively. Each of the two \ac{DRL} algorithms was trained for a total of $50 {\rm card}(\mathcal{A})$ time steps for each trial, followed by an evaluation period of $300$ steps, in which the agents selected actions with their learned deterministic policy (i.e., without choosing a random action with probability $\epsilon$). For fairness, identical neural networks are used by the two agents, although recall that \ac{DQN} uses two copies of its Q network. The employed network is consisted of two Convolutional/MaxPooling blocks followed by two fully connected layers with ReLU activations.
The convolutional layers have $64$ units and a kernel size of $5$, the MaxPooling operations also have a size of $5$, and the fully connected layers are consisted of $32$ units each, with the Dropout technique being applied for regularization.
The hyper-parameters used in the evaluation process are given in Table~\ref{tab:hyper-parameters}.
The \ac{UCB} algorithm was trained for $500 {\rm card}(\mathcal{A})$ steps to compensate for the lack of contextual observations.

\begin{table}[!t]
    \centering
    \caption{Parameter Values Used for the Simulation Results.}
    \vspace*{-5pt}
    \begin{tabular}{|l|c|}
        \hline
        \textbf{Parameter}                    & \textbf{Value} \\
        \hline
        \ac{BS} coordinates (m)               & $(10,5,2)$  \\
        \hline
        RIS coordinates (m)               & $(7.5,  13, 2)$, $(12.5, 13, 2)$    \\
        \hline
        \ac{UE}$_1$ coordinates (m)           & $(8.775, 14.394,  1.634)$    \\
        \hline
        \ac{UE}$_2$ coordinates (m)           & $(9.648, 13.281,  1.632)$    \\
        \hline
        $N_{\rm T}$, ${\rm card}(\mathcal{V})$                           & $4$ \\
        \hline
        $\kappa_1$, $\kappa_2$                & $30$ dB \\
        \hline
        $\sigma^2$ (equal for all \acp{UE})   & $-110$ dBm \\
        \hline
        Carrier frequency                     & $35$ GHz \\
        \hline
        $N_{\rm group}$                       & $16$    \\
        \hline
    \end{tabular}\vspace{-0.3cm}
    \label{tab:setup-parameters}
\end{table}

\begin{table}[!t]
    \centering
    \caption{Hyper-parameter values of the considered \ac{DRL} algorithms.}
    \vspace*{-5pt}
    \begin{tabular}{|l|c|}
        \hline
        \multicolumn{2}{|c|}{\textbf{Common Parameters}} \\
        \hline
        $\epsilon$          & 0.3 \\
        \hline
        Dropout probability & 0.2 \\
        \hline
        \multicolumn{2}{|c|}{\textbf{DRP Parameters}} \\
        \hline
        Learning rate $\eta_2$       & 0.001 \\
        \hline
        \multicolumn{2}{|c|}{\textbf{DQN Parameters}} \\
        \hline
        Batch size ${\rm card}(\mathcal{B})$               & $128$ \\
        \hline
        Learning rate $\eta_1$       & $0.0002$ \\
        \hline
        Soft update  $\tau$          & $0.18$ \\
        \hline
        Target update frequency $t'$ & $100$ \\
        \hline
    \end{tabular}\vspace{-0.1cm}
    \label{tab:hyper-parameters}
\end{table}

\begin{figure}[!t]
    \centering
    \includegraphics[width=0.8\linewidth]{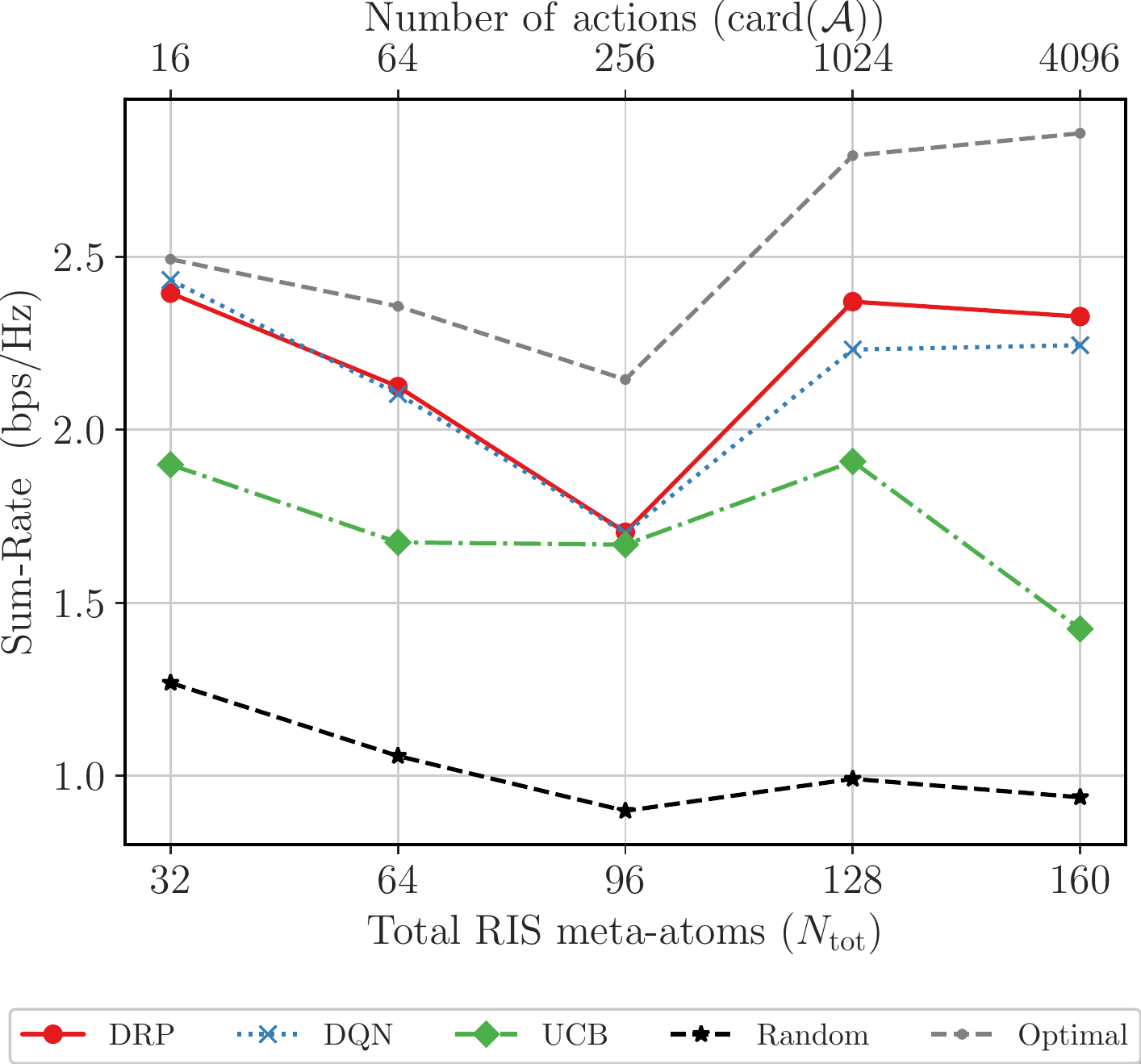}
    \vspace*{-7pt}
    \caption{Sum-rate performance of the compared algorithms as the number of RIS elements and the dimension of the action space increase. The BS transmit power $P$ was set to $40~{\rm dBm}$.}\vspace{-0.3cm}
    \label{fig:sum-rate-varying-N}
\end{figure}

Firstly, the average sum rates attained during the evaluation period by each method are compared in Fig.~\ref{fig:sum-rate-varying-N} across increasing \ac{RIS} sizes and different elements' groupings.
Clearly, the proposed \ac{DRP} algorithm and \ac{DQN} exhibit identical performances.
This result reinforces our hypothesis that elaborate \ac{MDP}-based techniques do not provide any significant advantage in the plain sum-rate maximization problem with \ac{IID} channel realizations. 
Both \ac{DRL} algorithms vastly outperform the random baseline, with an increase of higher than $89\%$.
At the same time, their performance is close to the optimal rate (varying approximately from $77\%$ to $96\%$).
Interestingly enough, the naive bandit approach, \ac{UCB}, is also capable of sufficiently outperforming the random selection strategy, while it achieves comparable results to the \ac{DRL} methods for the case of $96$ \ac{RIS} elements.
At the same time, Fig.~\ref{fig:sum-rate-varying-N} shows that its performance degrades at the largest trial and we expect the trend to continue in increasing \ac{RIS} sizes.
Nevertheless, it can be inferred that the formulation of the sum-rate maximization problem as a \ac{MAB} approach allows for channel-agnostic strategies to be deployed, albeit with relatively limited capabilities.

\begin{figure}[t]
    \centering
    \includegraphics[width=0.9\linewidth]{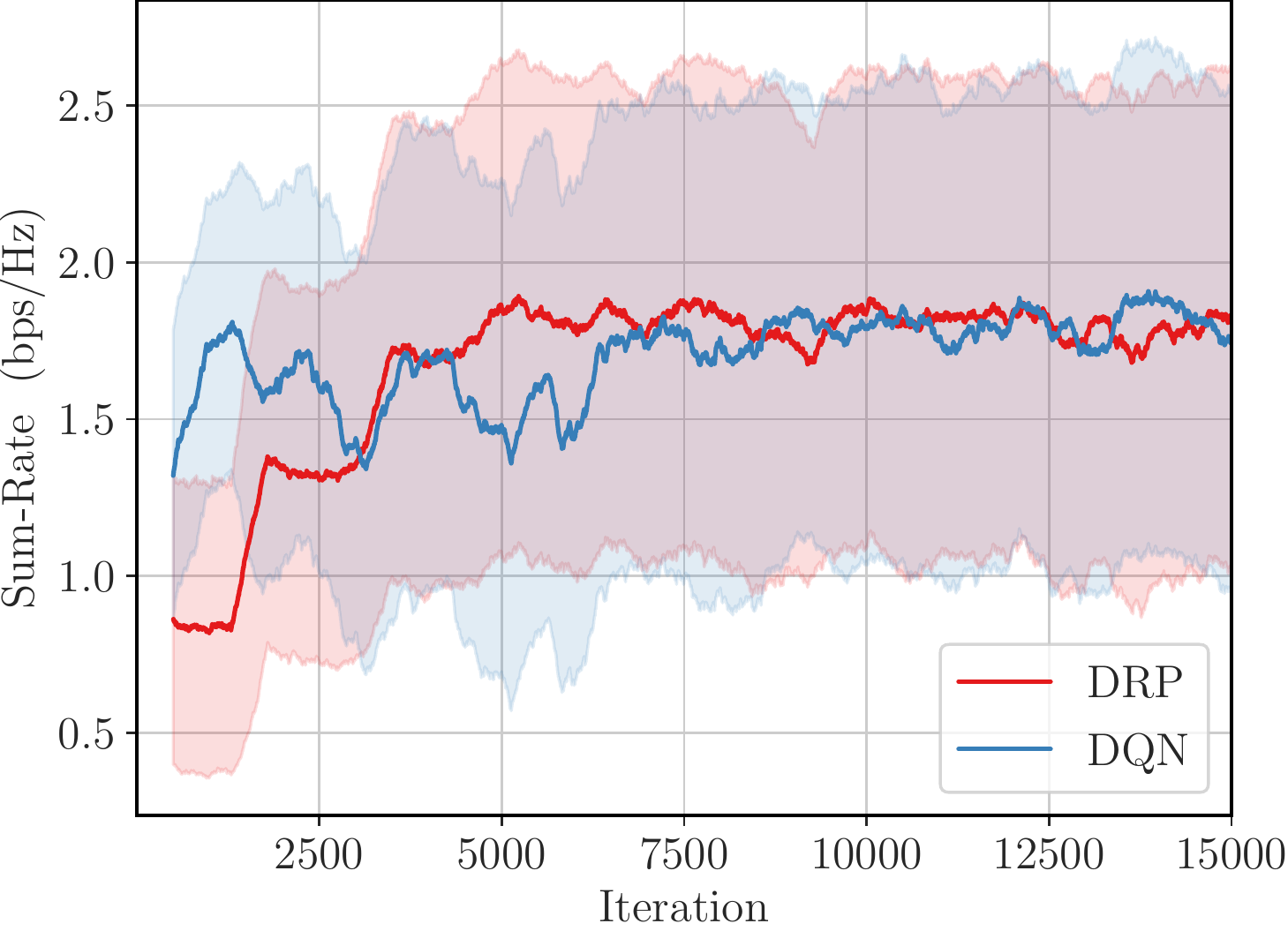}
    \vspace*{-7pt}
    \caption{Training curves of the compared DRL algorithms for the first portion of the training period. The curves are smoothed by averaging within a rolling window of $300$ iterations, and the shaded areas represent the accompanying standard deviations. $P$ is set to $40~{\rm dBm}$ and $N_{\rm tot}$ to $96$.}\vspace{-0.1cm}
    \label{fig:convergence-rate}
\end{figure}

In Fig.~\ref{fig:convergence-rate}, the attained rewards during the training process of the \ac{DRL} algorithms are depicted, for the setup with $128$ total \ac{RIS} elements.
For clarity, only the first $15000$ iterations are shown (approximately $30\%$ of the training period), in which variations during training are prominent.
It can be observed that the learning curve of the \ac{DQN} is steeper than that of \ac{DRP}, although they both reach their common peak plateau at approximately the same time.
Let it be noted that for this part of the evaluation, the actions are selected with the exploration policy (i.e. the $\epsilon$-greedy selection), which results in lower reward values, compared with the final evaluation of the learned (deterministic) policies.

To investigate the effect of the transmit power $P$ on the performance of the \ac{DRL} algorithms, we repeat the evaluation process for the trial with $96$ total \ac{RIS} elements, while varying $P$. The normalized average sum rates, with respect to the optimal rate given by exhaustive search, are given in Table~\ref{tab:varying-power}. The employed algorithms are mostly unaffected by the changes in $P$, performing in a similar manner (with respect to the optimal policy of exhaustive search), in all regimes. 

\begin{table}[!t]
    \centering
    \caption{Normalized Average Sum Rates for Varying Power.}
    \vspace*{-5pt}
    \begin{tabular}{|l|c|c|c|c|c|}
    \hline
	$P~{\rm (dBm)}$ & 10 & 20 & 30 & 40 & 50 \\
	\hline  
	DRP  & 0.728 & 0.791 & 0.787 & 0.800 & 0.746 \\
	\hline
	DQN & 0.731 & 0.754 & 0.775 & 0.809 & 0.737 \\
	\hline
    \end{tabular}\vspace{-0.1cm}
    \label{tab:varying-power}
\end{table}

%% file: Sections/6_Conclusion.tex
\vspace{-0.3cm}
\section{Conclusion}\label{sec:Conclusion}
\vspace{-0.1cm}
This paper addressed the sum-rate maximization problem in \ac{RIS}-empowered multi-user MISO systems using \ac{RL} techniques.
Motivated by the \ac{IID} assumption about the channel realizations and the immediate rate feedback, we suggested a treatment of the problem under the \ac{MAB} framework, instead of the traditional \ac{MDP} formalism.
We proposed a \ac{DRL} bandit algorithm equipped with a reward prediction network to estimate average sum rates per action (RIS configuration and precoder selection) and the $\epsilon$-greedy exploration behavior.
The numerical evaluation process in a multi-RIS system established that the method compares equally with the state-of-the-art \ac{DQN} algorithm, while being simpler in terms of interpretability, number of hyper-parameters, and neural network requirements.
Finally, we demonstrated that our MAB-based \ac{UCB} approach attains competitive performance in certain cases without the need for any channel observation.  

%% file: main.bbl
\begin{thebibliography}{10}
\providecommand{\url}[1]{#1}
\csname url@samestyle\endcsname
\providecommand{\newblock}{\relax}
\providecommand{\bibinfo}[2]{#2}
\providecommand{\BIBentrySTDinterwordspacing}{\spaceskip=0pt\relax}
\providecommand{\BIBentryALTinterwordstretchfactor}{4}
\providecommand{\BIBentryALTinterwordspacing}{\spaceskip=\fontdimen2\font plus
\BIBentryALTinterwordstretchfactor\fontdimen3\font minus
  \fontdimen4\font\relax}
\providecommand{\BIBforeignlanguage}[2]{{%
\expandafter\ifx\csname l@#1\endcsname\relax
\typeout{** WARNING: IEEEtran.bst: No hyphenation pattern has been}%
\typeout{** loaded for the language `#1'. Using the pattern for}%
\typeout{** the default language instead.}%
\else
\language=\csname l@#1\endcsname
\fi
#2}}
\providecommand{\BIBdecl}{\relax}
\BIBdecl

\bibitem{Saad_6G_2020}
W.~Saad \emph{et~al.}, ``A vision of {6G} wireless systems: {A}pplications,
  trends, technologies, and open research problems,'' \emph{IEEE Network},
  vol.~34, no.~3, pp. 134--142, Jun. 2020.

\bibitem{EURASIP_RIS_paper}
M.~Di~Renzo \emph{et~al.}, ``Smart radio environments empowered by
  reconfigurable {AI} meta-surfaces: An idea whose time has come,''
  \emph{EURASIP J. Wireless Commun. Netw.}, vol. 2019, no.~1, pp. 1--20, May
  2019.

\bibitem{LZAWFM2020}
S.~Lin \emph{et~al.}, ``Adaptive transmission for reconfigurable intelligent
  surface-assisted {OFDM} wireless communications,'' \emph{IEEE J. Sel. Areas
  Commun.}, vol.~38, no.~11, pp. 2653--2665, Nov. 2020.

\bibitem{alexandg_2021}
G.~C. Alexandropoulos \emph{et~al.}, ``Reconfigurable intelligent surfaces for
  rich scattering wireless communications: {R}ecent experiments, challenges,
  and opportunities,'' \emph{IEEE Commun. Mag.}, vol.~59, no.~6, pp. 28--34,
  Jun. 2021.

\bibitem{rise6g}
E.~Calvanese~Strinati \emph{et~al.}, ``Wireless environment as a service
  enabled by reconfigurable intelligent surfaces: {T}he {RISE-6G}
  perspective,'' in \emph{Proc. Joint EuCNC \& 6G Summit}, Porto, Portugal,
  Jun. 2021.

\bibitem{huang2019reconfigurable}
C.~Huang \emph{et~al.}, ``Reconfigurable intelligent surfaces for energy
  efficiency in wireless communication,'' \emph{IEEE Trans. Wireless Commun.},
  vol.~18, no.~8, pp. 4157--4170, Aug. 2019.

\bibitem{5GPPP_AIML}
``{AI} and {ML} – {E}nablers for beyond {5G} networks,'' White Paper, 5G PPP
  Technology Board, May 2021.

\bibitem{AIRIS}
S.~Zhang \emph{et~al.}, ``{AIRIS}: {Artificial} intelligence enhanced signal
  processing in reconfigurable intelligent surface communications,''
  \emph{China Commun.}, vol.~18, no.~7, pp. 158--171, 2021.

\bibitem{huang2019spawc_indoor}
C.~{Huang} \emph{et~al.}, ``Indoor signal focusing with deep learning designed
  reconfigurable intelligent surfaces,'' in \emph{Proc. IEEE SPAWC}, Cannes,
  France, Jul. 2019.

\bibitem{Abdelrahman2020TowardsStandaloneOperation}
A.~Taha \emph{et~al.}, ``Deep reinforcement learning for intelligent reflecting
  surfaces: Towards standalone operation,'' in \emph{Proc. IEEE SPAWC},
  Atlanta, USA, May 2020.

\bibitem{Huang2020DRL_RIS}
C.~Huang \emph{et~al.}, ``Reconfigurable intelligent surface assisted multiuser
  {MISO} systems exploiting deep reinforcement learning,'' \emph{IEEE J. Sel.
  Areas Commun.}, vol.~38, no.~8, pp. 1839--1850, Aug. 2020.

\bibitem{Feng2020DRL_MISO}
K.~Feng \emph{et~al.}, ``Deep reinforcement learning based intelligent
  reflecting surface optimization for {MISO} communication systems,''
  \emph{IEEE Wireless Commun. Lett.}, vol.~9, no.~5, pp. 745--749, May 2020.

\bibitem{Huang2021MultiHop}
C.~Huang \emph{et~al.}, ``Multi-hop {RIS}-empowered terahertz communications:
  {A DRL}-based hybrid beamforming design,'' \emph{IEEE J. Sel. Areas Commun.},
  vol.~39, no.~6, pp. 1663--1677, Jun. 2021.

\bibitem{kim2021multiirsassisted}
J.~Kim \emph{et~al.}, ``Multi-{IRS}-assisted multi-cell uplink {MIMO}
  communications under imperfect {CSI}: {A} deep reinforcement learning
  approach,'' 2021, [Online] https://arxiv.org/pdf/2011.01141.pdf.

\bibitem{Liu2021MNOMA_deployment}
X.~Liu \emph{et~al.}, ``{RIS} enhanced massive non-orthogonal multiple access
  networks: {D}eployment and passive beamforming design,'' \emph{IEEE J. Sel.
  Areas Commun.}, vol.~39, no.~4, pp. 1057--1071, Apr. 2021.

\bibitem{Samir2021AgeOfInformation}
M.~Samir \emph{et~al.}, ``Optimizing age of information through aerial
  reconfigurable intelligent surfaces: A deep reinforcement learning
  approach,'' \emph{IEEE Trans. Veh. Technol.}, vol.~70, no.~4, pp. 3978--3983,
  Apr. 2021.

\bibitem{Yang2021DRL_for_Secure}
H.~Yang \emph{et~al.}, ``Deep reinforcement learning-based intelligent
  reflecting surface for secure wireless communications,'' \emph{IEEE Trans.
  Wireless Commun.}, vol.~20, no.~1, pp. 375--388, Jan. 2021.

\bibitem{Lee2020DRL_EE}
G.~Lee \emph{et~al.}, ``Deep reinforcement learning for energy-efficient
  networking with reconfigurable intelligent surfaces,'' in \emph{Proc. IEEE
  ICC}, Dublin, Ireland, Jun. 2020.

\bibitem{gao2021ResourceAllocation}
X.~Gao \emph{et~al.}, ``Machine learning empowered resource allocation in {IRS}
  aided {MISO-NOMA} networks,'' 2021, [Online]
  https://arxiv.org/pdf/2103.11791.pdf.

\bibitem{alhilo2021reconfigurable}
A.~Al-Hilo \emph{et~al.}, ``Reconfigurable intelligent surface enabled
  vehicular communication: Joint user scheduling and passive beamforming,''
  2021, [Online] https://arxiv.org/pdf/2101.12247.pdf.

\bibitem{Alkhateeb_JSTSP_all}
A.~Alkhateeb \emph{et~al.}, ``Channel estimation and hybrid precoding for
  millimeter wave cellular systems,'' \emph{{IEEE} {J.} {S}el. {T}opics
  {S}ignal {P}rocess.}, vol.~8, no.~5, pp. 831--846, Oct. 2014.

\bibitem{Lin2021}
S.~Lin \emph{et~al.}, ``{Reconfigurable intelligent surfaces with reflection
  pattern modulation: Beamforming design, channel estimation, and achievable
  rate analysis},'' \emph{{IEEE Trans. Wireless Commun.}}, vol.~20, no.~2, pp.
  741--754, Feb. 2021.

\bibitem{Alamzadeh2021ris}
I.~Alamzadeh \emph{et~al.}, ``{A reconfigurable intelligent surface with
  integrated sensing capability},'' \emph{Scientific Reports}, vol.~11, no.~1,
  p. 20737, 2021.

\bibitem{DQN}
V.~Mnih \emph{et~al.}, ``Human-level control through deep reinforcement
  learning,'' \emph{Nature}, vol. 518, no. 7540, pp. 529--533, 2015.

\bibitem{UCB}
R.~Agrawal, ``Sample mean based index policies with {$O(\log n)$} regret for
  the multi-armed bandit problem,'' \emph{Adv. Applied Probability}, vol.~27,
  no.~4, pp. 1054--1078, 1995.

\end{thebibliography}
